\documentclass[reprint, nobibnotes, aps]{revtex4-1}

\usepackage{graphicx}% Include figure files
\usepackage{dcolumn}% Align table columns on decimal point
\usepackage{bm}% bold math
\usepackage[dvipdfm,colorlinks=true,linkcolor=blue,citecolor=red,urlcolor=blue]{hyperref}% add hypertext capabilities
\usepackage{mathrsfs,mathtools,color,wasysym,dsfont,here}
\usepackage{physics}
\usepackage[all]{xy}
\usepackage{tikz}
\usepackage{amsmath, amssymb, amsfonts}
\usetikzlibrary{calc}
\usetikzlibrary{arrows}
\usetikzlibrary{decorations.markings,decorations.pathmorphing}

\begin{document}

%\preprint{APS/123-QED}

\title{Dissipative spin chain as a non-Hermitian Kitaev ladder}% Force line breaks with \\

\author{Naoyuki Shibata}
\email{shibata-naoyuki@g.ecc.u-tokyo.ac.jp}
\author{Hosho Katsura}%
\affiliation{%
 Department of Physics, Graduate School of Science, the University of Tokyo, Hongo, Tokyo 113-0033, Japan
}%

%\affiliation{
%\affiliation{
%\affiliation{%

\date{\today}% It is always \today, today,

\begin{abstract}
	We derive exact results for the Lindblad equation for a quantum spin chain (one-dimensional quantum compass model) with dephasing noise. The system possesses doubly degenerate nonequilibrium steady states due to the presence of a conserved charge commuting with the Hamiltonian and Lindblad operators. 
We show that the system can be mapped to a non-Hermitian Kitaev model on a two-leg ladder, which is solvable by representing the spins in terms of Majorana fermions. This allows us to study the Liouvillian gap, the inverse of relaxation time, in detail. We find that the Liouvillian gap increases monotonically when the dissipation strength $ \gamma $ is small, while it decreases monotonically for large $ \gamma $, implying a kind of phase transition in the first decay mode. The Liouvillian gap and the transition point are obtained in closed form in the case where the spin chain is critical. We also obtain the explicit expression for the autocorrelator of the edge spin. The result implies the suppression of decoherence when the spin chain is in the topological regime.
%\begin{description}
%\end{description}
\end{abstract}

\pacs{Valid PACS appear here}% PACS, the Physics and Astronomy
\maketitle

%\tableofcontents

\section{Introduction}%
	With recent advances in quantum engineering, it becomes increasingly important to study how the coupling to the environment affects a system. The time evolution of such an open system can be described by a master equation. %With 
Under rather general conditions that the evolution is Markovian and completely positive and trace preserving (CPTP), one obtains the Lindblad equation \cite{Breuer2002} for the time-dependent density matrix. In the past, this quantum master equation %had been applied in the past
had been mostly used to describe few-particle systems in, e.g., quantum optics. %recently 
However, recent years have witnessed a growing interest in many-particle systems in the Lindblad setting %has been studied
\cite{Prosen2009,Znidaric2014,Znidaric2015,Medvedyeva2016}. %intensively.
	
	Although there are several approaches to analyze the Lindblad equation %including the using 
such as perturbation theory \cite{Gallis1996,Li2014} and numerical %calculation 
methods \cite{Prosen2009,Cui2015,Kshetrimayum2018,Raghunandan2018}, 
%it can be more appreciated to obtain exact solutions. 
exact results for the full dynamics is few and far between. 
In some cases \cite{Prosen2008,Prosen2011}, the nonequilibrium steady states (NESSs) can be constructed exactly, but it is more challenging to %solve 
completely %namely, to 
diagonalize the Liouvillian (the generator of the Lindblad equation). 
In this sense, much fewer cases are known as exactly solvable models~\cite{Prosen2008,Medvedyeva2016}. 
The difficulty %to analyze the Lindblad equation is due to the fact that we have to deal
lies in dealing with the space of linear operators, the dimension of %this space 
which grows more rapidly than that of the Hilbert space. 
This limits the system size %at which numerical exact diagonalization method can use. 
amenable to exact numerical diagonalization. 
To make matters worse, it is often the case that effective interactions arise from dissipation even when the Hamiltonian itself is reducible to that of a free-particle system. This prevents us from understanding the full dynamics of the system.
	
In this paper, we present an exactly solvable dissipative model which corresponds to the non-Hermitian many-body quantum system. 
%Our model will have 
%We expect that there are several ways to simulate our model using, e.g., trapped atomic ions \cite{Jurcevic2014,Richerme2014} or ultracold atoms in optical lattices \cite{Duan2003}. 
%%Dephasing noise we consider here as dissipation kills \textbf{the} interference \textbf{between (?) the} off-diagonal terms, hence, destroys the quantum coherence. 
%Dephasing noise we consider here as dissipation kills the off-diagonal elements of the density matrix. In other words, the system loses its quantum %interference property, namely, 
%coherence. 
%It is a challenging task for quantum engineering how to suppress the decoherence caused by the environment. 
The model we propose has a conserved charge that leads to two exact NESSs. Moreover, our model can be seen as a non-Hermitian Kitaev model on a two-leg ladder \cite{DeGottardi2011,Wu2012}. Therefore, by applying Kitaev's technique \cite{Kitaev2006}, our model can be mapped to free Majorana fermions in a static $ \mathbb{Z}_2 $ gauge field, 
%hence, the Liouvillian is fully diagonalized. 
which allows us to fully diagonalize the Liouvillian. 
%The numerical calculation suggests
We numerically identify the gauge sectors where the first decay modes live. Then, %under the assumption of numerically obtained gauge flux configuration, 
assuming the flux configurations obtained, we derive the exact Liouvillian gap, the inverse of the relaxation time. 
We also study the infinite temperature autocorrelator of an edge spin and obtain its exact formula by applying techniques from combinatorics. 
		
\section{Models and NESS\lowercase{s}}
	We consider the Lindblad equation for the density matrix $\rho$
	\begin{align}
		\dv{\rho}{t}=\mathcal{L}[\rho]\coloneqq-\mathrm{i}[H,\rho]+\sum_{i}\qty(L_i\rho L_i^\dagger-\dfrac{1}{2}\qty{L_i^\dagger L_i,\rho}),
   \label{eq:Lindblad}
	\end{align}
where $H$ denotes the Hamiltonian of the one-dimensional quantum compass model \cite{Brzezicki2007,Feng2007,You2008,Eriksson2009,Jafari2011,Liu2012} given by
	\begin{align}
		H=-\sum_{i=1}^{N/2} J_x\sigma_{2i-1}^x\sigma_{2i}^x -\sum_{i=1}^{N/2-1} J_y\sigma_{2i}^y\sigma_{2i+1}^y,\label{eq:alternate_XY_chain}
	\end{align}
	and $ L_i=\sqrt{\gamma}\sigma_i^z\, (i=1,\dots,N) $ are Lindblad operators. This form of dissipation is known as dephasing noise~\cite{Cai2013,Znidaric2015,VanCaspel2018} which kills off-diagonal elements of the density matrix, and hence destroys the quantum coherence. Here,
%$ \sigma^x, \sigma^y, \sigma^z $ are the standard Pauli matrices, 
$ \sigma^\alpha_j $ ($\alpha=x,y,z$) are the Pauli operators at site $j$, $ J_x $ and $ J_y $ are the exchange couplings (in the unit of energy), $ \gamma\ge0 $ is the dissipation strength parameter, and $ N $ is the number of site. We assume %the open boundary condition and 
	$ N $ is even and the open boundary conditions are imposed.
It suffices to consider the case $ J_x,J_y\ge 0 $, as the other cases can be obtained by an appropriate unitary transformation. 
The operator $ \mathcal{L}[\rho] $ is called a Liouvillian or a Lindbladian. A NESS is a fixed point of the dynamics Eq. (\ref{eq:Lindblad}), i.e., an eigenstate of the Liouvillian $ \mathcal{L} $ with eigenvalue $ 0 $. 
%$ \mathcal{L}[\rho_\infty]=0 $.
For our model, there are two steady states $ \rho_\pm\coloneqq (\mathds{1}\pm Q)/2^N $, where $ \mathds{1} $ is an identity matrix and $ Q\coloneqq\prod_{i} \sigma_i^z $ %, that can be shown as follows. Note that $ Q $ 
	is a conserved charge, i.e.,
	\begin{align}
	\qty[H, Q]=0,\; \qty[L_i,Q]=\qty[L_i^\dagger ,Q]=0\quad\text{for }\forall i.
	\end{align}

	The proof goes as follows. Because of the Hermiticity of Lindblad operators, there is a trivial NESS, i.e., \emph{completely mixed state} $ \rho_\mathrm{c}\coloneqq \mathds{1}/2^N $. One can easily verify that if $ \mathcal{L}[\rho_\mathrm{c}]=0 $, then $ \rho = Q\rho_\mathrm{c} $ also satisfies $ \mathcal{L}[\rho]=0 $. 
Although $ \rho $ itself is not positive semi-definite which is a necessary condition to be a density operator, one can construct the following operators $ P_{\pm}\coloneqq (\mathds{1}\pm Q)/2 $, which are orthogonal projections, and hence positive semi-definite. %with a linear combination 
	This then gives (normalized) steady states $ \rho_\pm = P_\pm /\tr P_\pm $. We have checked numerically for small $N$ that they are the unique NESS of the system. 
	%the system has exactly two steady states. 

\section{Mapping to Kitaev ladder}
\begin{figure}
		\centering
		\includegraphics[width=1.0\linewidth]{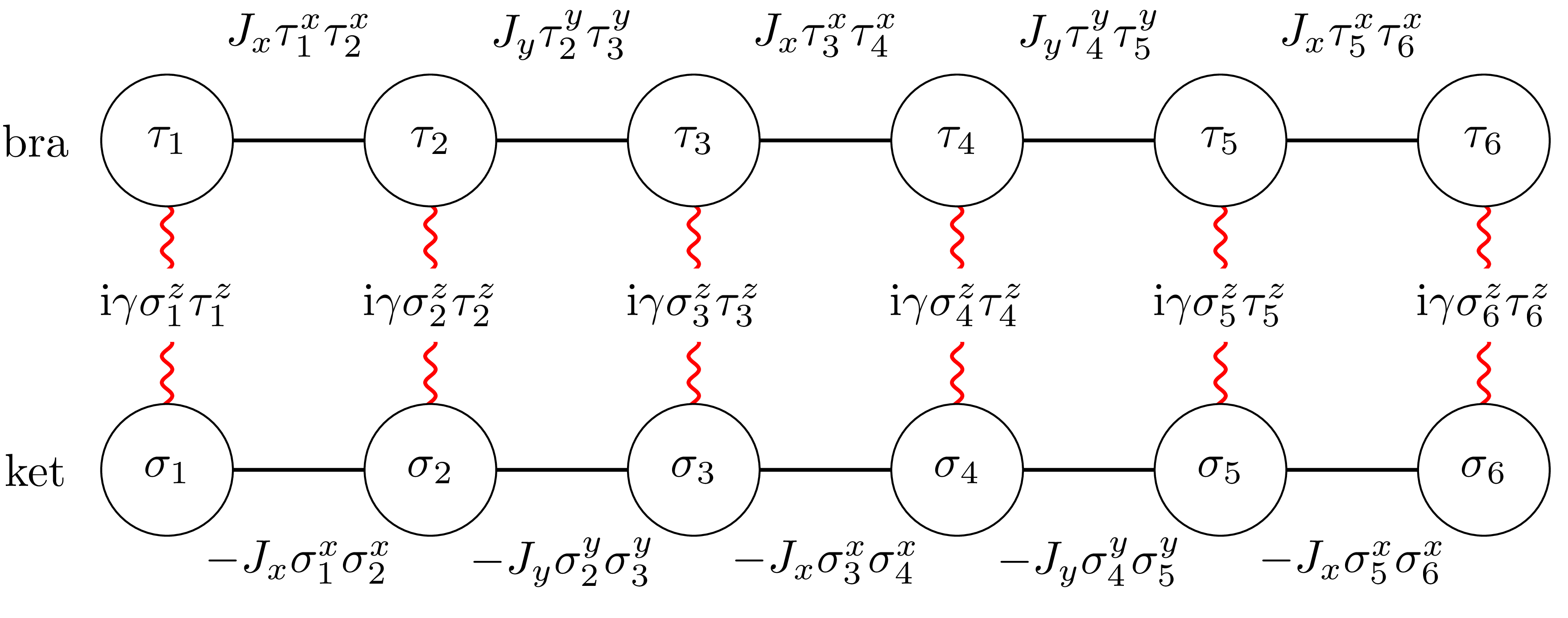}
		\caption{The non-Hermitian ladder system for the one-dimensional quantum compass model with dephasing noise. By Kitaev's mapping, this can be seen as a model of free Majorana fermions in a static $ \mathbb{Z}_2 $ gauge field. We fix all the signs of the links except for those of red wavy lines (see Appendix \ref{app:from_Kitaev_to_SSH} for more details).}
		\label{fig:lindblad-kitaev-ladder}
	\end{figure}

	A $ 2^N\times 2^N $ density matrix $ \rho $ can be thought of as a $ 2^{2N}\ $-dimensional vector (see Appendix \ref{app:Lindbladian_non-Hermitian} for details). In this sense, we can identify the Liouvillian $ \mathcal{L} $ for the one-dimensional chain as a non-Hermitian Hamiltonian of a ladder system \cite{Znidaric2014,Znidaric2015} (see Fig. \ref{fig:lindblad-kitaev-ladder}):
	\begin{equation}
		\mathrm{i}(\mathcal{L}+\gamma N)\cong H\otimes\mathds{1}-\mathds{1}\otimes H+\sum_{i=1}^N \mathrm{i}\gamma\sigma_i^z\otimes\tau_i^z\eqqcolon\mathcal{H},\label{eq:XY_ladder_Hamiltonian}
	\end{equation}
	where the Hilbert space of the RHS is the ``$ \mathrm{Ket}\otimes\mathrm{Bra} $ space'' and $ \tau_i^z $ is the Pauli matrix for the $ i $th Bra site. The non-unitary terms in the Liouvillian (\ref{eq:Lindblad}) correspond to the non-Hermitian terms in $ \mathcal{H} $. 
	
	We apply to this ladder Hamiltonian (\ref{eq:XY_ladder_Hamiltonian}) the technique by Kitaev \cite{Kitaev2006}, which was originally used to solve the quantum spin model on a honeycomb lattice. 
	In order to solve the model, Kitaev developed an elegant technique: substituting \emph{Majorana fermion operators} for spin operators $ \sigma_j^\alpha\to \mathrm{i}b_j^\alpha c_j $. Here, $ b_j^\alpha $ and $ c_j $ are Majorana operators obeying the Clifford algebra $ \{c_j, c_k\}=2\delta_{jk},\, \{b_j^\alpha, b_k^\beta\}=2\delta_{jk}\delta^{\alpha\beta}$, and $ \{b_j^\alpha, c_k\}=0 $ with $ \qty{A,B}=AB+BA $ being the anti-commutator. After the mapping, we have a quadratic Hamiltonian of itinerant Majorana fermions %$ \qty{c_i}_i $ 
($c_i$'s) in each sector specified by the static $ \mathbb{Z}_2 $ gauge field %consisting of $ \qty{b_i^\alpha}_{i,\alpha} $ 
	(i.e., each link has the sign $ \pm 1 $ in the hopping %coefficient 
	amplitude). 
	Thus, we can diagonalize the Hamiltonian and obtain all eigenvalues and eigenstates, sector by sector. Besides the honeycomb lattice, Kitaev's mapping is applicable to other lattices with a similar Hamiltonian.
	% including the ladder system 
	Examples include the ladder system \cite{DeGottardi2011}, which is the Hermitian analog of our model (\ref{eq:XY_ladder_Hamiltonian}).
	
	Next, we define complex fermions $ f_i,f_i^\dagger $ each of which is made of two Majorana fermions $ c_i $ and $ d_i $: $ f_i\coloneqq (c_i+\mathrm{i}d_i)/2,\quad f_i^\dagger\coloneqq (c_i-\mathrm{i}d_i)/2 $. 
	Here, $ c_i $ (respectively, $ d_i $) is the Majorana operator for the $ \sigma_i $ (respectively, $ \tau_i $) spin. 
Then,
	%after the unitary transformation $ f_j\to \mathrm{e}^{\mathrm{i}(\pi/2)j}f_j $, 
	the model is mapped to the Su-Schrieffer-Heeger (SSH) model \cite{Su1979} with imaginary chemical potential \cite{Klett2017,Lieu2018}. The Hamiltonian reads (see Appendix \ref{app:from_Kitaev_to_SSH} for details)
	\begin{align}
		\mathcal{H}(\vb*{\mu})&=
		-\sum_{i=1}^{N}\mathrm{i}\gamma\mu_i
		+\sum_{k,l} \mathsf{A}_{kl} f_k^\dagger f_l,
		\label{eq:non-Hermitian_SSH_1}
	\end{align}
	where $ \vb*{\mu}=(\mu_1,\dots, \mu_N) $ and $ \mathsf{A} $ is a tridiagonal and complex-symmetric matrix given by
	\begin{align}
		\mathsf{A}&\coloneqq 2
		\begin{pmatrix}
			\mathrm{i}\gamma\mu_1& J_x&&&\\
			J_x&\mathrm{i}\gamma \mu_2&J_y&&\\
			&J_y&\mathrm{i}\gamma\mu_3&\ddots&&\\
			&&\ddots&\ddots&J_x\\
			&&&J_x&\mathrm{i}\gamma \mu_{N}
		\end{pmatrix}.
	\end{align}
	$ \mu_i $'s come from the gauge degree of freedom and take the value $ \pm 1 $.
	The solution of a non-Hermitian quadratic form of fermions is similar to that of a %usual 
	Hermitian one. 
One can construct many-body eigenstates of $\mathcal{H}(\vb*{\mu})$ by just filling single-particle energies of the Hamiltonian, which can be obtained by diagonalizing $ \mathsf{A} $.

Symmetries of the Hamiltonian (\ref{eq:non-Hermitian_SSH_1}) enable us to restrict the configurations of $\mu_i$'s to consider. 
First, $ \mathcal{H}\qty(\vb*{\mu}) $ and $ \mathcal{H}\qty(-\vb*{\mu}) $ have the same spectrum because 
%the transformation $ \vb*{\mu}\to-\vb*{\mu} $ does not change the flux phase in the ladder system. 
the flux configuration in the ladder system is invariant under sending $ \vb*{\mu}\to-\vb*{\mu} $. 
(In view of the SSH model, $ \mathcal{H}\qty(\vb*{\mu}) $ is transformed to $ \mathcal{H}\qty(-\vb*{\mu}) $ by the charge conjugation $ f_j\to(-1)^j f_j^\dagger $.) Second, due to the inversion symmetry, the transformation $ (\mu_1,\mu_2,\dots,\mu_N)\to(\mu_N,\mu_{N-1},\dots,\mu_1) $ leaves the spectrum of $ \mathcal{H} $ unchanged. 
In the following, we only consider the configurations in which the number of positive $\mu_i$'s is not less than the number of negative $\mu_i$'s. 

\section{Liouvillian gap}%
	\begin{figure*}
		\centering
		\includegraphics[width=1.0\linewidth]{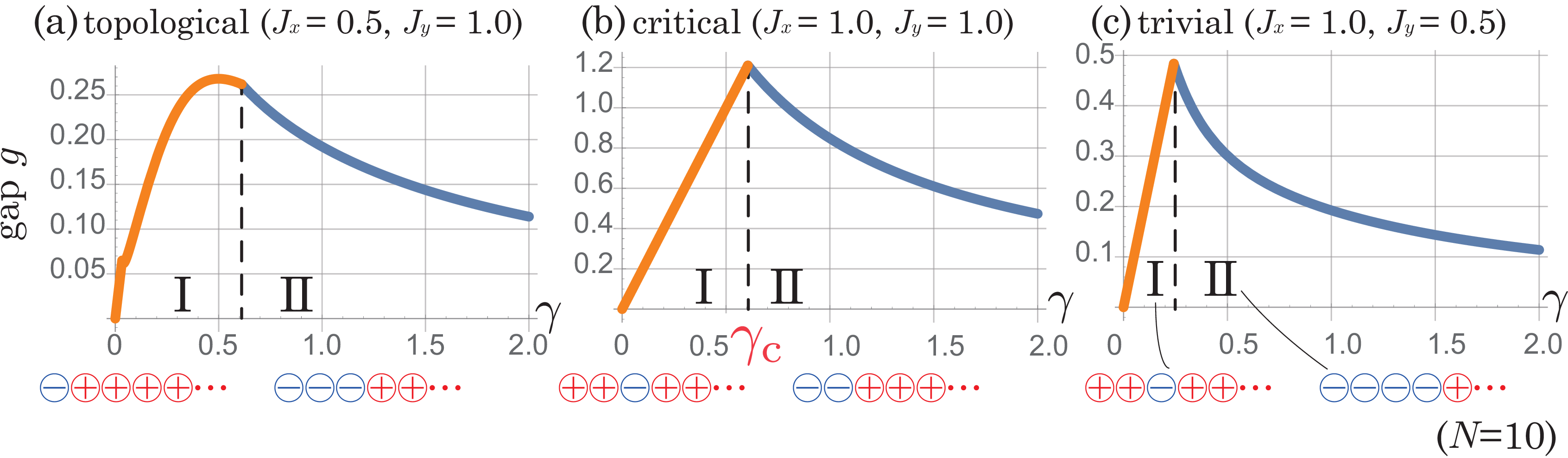}
		\caption{(Color online) Numerical results of the Liouvillian gap $ g $ as a function of the dissipation strength $ \gamma $. (a) Topological regime ($ J_x = 0.5$, $J_y=1.0$), (b) critical point ($ J_x=J_y=1.0 $), and (c) trivial regime ($ J_x=1.0 $, $J_y=0.5$). 
%All cases have a cusp indicated by the dashed line. 
The position of the cusp is indicated by the dashed line in each case. 
%One example of the chemical potential configuration $ \vb*{\mu} $ where the first decay mode lives is also shown. 
The $+/-$ pattern below each phase shows a chemical-potential configuration $ \vb*{\mu} $ that gives the first decay mode. }
		\label{fig:lindblad-kitaevgvsgap}
	\end{figure*}Let eigenvalues of the Liouvillian $ \mathcal{L} $ be $ \Lambda_i(\mathcal{L}) $. It can be proved \cite{Breuer2002,Rivas2011} that all $ \Lambda_i(\mathcal{L}) $ satisfy $ \Re[\Lambda_i(\mathcal{L})] \le 0 $. A Liouvillian gap $ g $ is defined as
	\begin{equation}
		g\coloneqq -\max_{\substack{i\\ \Re[\Lambda_i(\mathcal{L})]\ne 0}}\Re[\Lambda_i(\mathcal{L})],\label{eq:def_of_Liouvillian_gap}
	\end{equation}
	hence, the inverse of the relaxation time. %As seen 
	It is clear from Eqs. (\ref{eq:XY_ladder_Hamiltonian}) and (\ref{eq:def_of_Liouvillian_gap}) that the Liouvillian gap corresponds to the gap between the first and second largest \textit{imaginary} parts of eigenvalues of $ \mathcal{H} $, the former of which is $ \gamma N $. The configuration which gives the eigenvalue $ \mathrm{i}\gamma N $ is $ \mu_{i}=+1 $ for all $ i $. 
	%that is based on the following reason. 
	The reason is as follows. 
	In this configuration, the one-particle energy levels are obtained just by shifting those of the original SSH model by $ +2\mathrm{i}\gamma $. Therefore, we obtain the eigenvalue $ \mathrm{i}\gamma N $ by filling all the energy levels. Then, the Liouvillian gap is recast as
	\begin{align}
		g&=\gamma N-\max_{\substack{\vb*{\mu}\\ m(\vb*{\mu})\ne 0}}\qty[\sum_{i=1}^{N}\qty(\dfrac{\Im\lambda_i+\abs{\Im\lambda_i}}{2}-\gamma \mu_{i})]\nonumber\\
		&=\min_{\substack{\vb*{\mu}\\ m(\vb*{\mu})\ne 0}}\qty(2m(\vb*{\mu})\gamma-\sum_{i=1}^{N}\dfrac{\abs{\Im\lambda_i}-\Im\lambda_i}{2}),\label{eq:Liouvillian_gap_simplified}
	\end{align}
	where $ \lambda_i $ denotes the $i$th eigenvalue of $ \mathsf{A} $, and $ m(\vb*{\mu}) $ the number of $ \mu_i $'s which are $ -1 $.
	%		(resp. $ {\sum}_< $) denotes the sum over all positive (resp. negative) arguments. 
Since at least one of $\mu_i$'s must be $ -1 $ when $ m(\vb*{\mu}) \ne 0 $, we have $g \le 2\gamma$. 
%One might think that we also need to consider removing fermions from single-particle energy levels in ``$ \forall i,\, \mu_i=+1 $ configuration''. 
%However, removing one fermion from its occupied state of this gauge configuration decreases the total energy by $ 2\mathrm{i}\gamma $. 
One might think that we need to consider the case where $ \mu_i=+1 $ for all $i$ and some single-particle energy levels are empty. However, this is not the case. The Liouvillian gap in this sector must be greater than or equal to $ g $ in Eq. (\ref{eq:Liouvillian_gap_simplified}), as removing one
fermion from an occupied state in this case decreases the imaginary part of the eigenvalue of $ {\cal H} $ by $ 2\gamma $.
Thus, it suffices to consider configurations different from the one with $\mu_i=+1$ for all $i$.

	Figure \ref{fig:lindblad-kitaevgvsgap} shows the numerical results of $ g $ as a function of $ \gamma $ for various $ J_y/J_x $ for a system size $ N=10 $. Here, 
%we use the word 
``topological,'' ``critical,'' and ``trivial'' cases refer to the regions $ J_y/J_x>1 $, $ J_y=J_x $, and $ J_y/J_x<1 $, respectively, in analogy with the 
%usual 
Hermitian SSH model. 
	From this figure, we can see a kind of phase transition of the first decay mode in every (topological, critical, or trivial) case. We also numerically obtained the chemical-potential configurations which give the first decay mode, as also shown in Fig. \ref{fig:lindblad-kitaevgvsgap}. We do not show all the configurations which give the same eigenvalues. In Appendix \ref{app:the_first_decay_modes_configurations}, up to symmetries mentioned above, every configuration which gives the first decay mode is shown.
In the ``phase I'' of the critical and %topological 
trivial cases, the gap behaves as exactly $ g=2\gamma $. The reason for this behavior becomes clear in Eq. (\ref{eq:Liouvillian_gap_simplified}). When $ \gamma $ is small enough, all $ \lambda_i $ satisfy $ \Im\lambda_i>0 $, then $ g=2\gamma $ follows. In the topological case, the situation is %a little 
slightly different. In this case there exists a $ \lambda_i $ which satisfies $ \Im\lambda_i<0 $, and $ g $ is smaller than $ 2\gamma $. For finite $ N $, there is another configuration in the region of $ \gamma \ll 1 $ (see Appendix \ref{app:the_first_decay_modes_configurations}), but this region shrinks to zero as $ N\to\infty $. In the ``phase I\hspace{-0.1em}I'', the gap behaves asymptotically $ g\propto 1/\gamma $ in each case. This increase in relaxation time as $ \gamma\to\infty $ can be thought of as the Quantum Zeno effect \cite{Vasiloiu2018}.

	Our extensive numerical calculation suggests that the chemical-potential configurations which give the first decay modes do not depend on the system size. Then, under this assumption, we can obtain the exact formula for the Liouvillian gap $ g $ and the transition point $ \gamma_\mathrm{c} $ in the thermodynamic limit of the critical case $ J_x=J_y=1 $ (see Appendix \ref{app:derivation_of_Liouvillian_gap} for more details):
	\begin{gather}
		g=
		\left\{
		\begin{array}{cc}
			2\gamma & \qty(0\le \gamma\le\gamma_\mathrm{c})\\
			\dfrac{6^{1/3}\qty(9 \gamma^2+\sqrt{48 \gamma^6+81\gamma^4})^{2/3}-2\cdot 6^{2/3} \gamma^2}{3 \gamma(9 \gamma^2+\sqrt{48 \gamma^6+81 \gamma^4})^{1/3}} & \qty(\gamma_\mathrm{c}\le\gamma)
		\end{array}
		\right. ,
     \label{eq:exact_Liouvillian_gap}
	\end{gather}
where
    \begin{align}
		  \quad\gamma_\mathrm{c}=\sqrt{\dfrac{\sqrt{3}-1}{2}}\simeq 0.605.
    \end{align}
We have confirmed that this result agrees well with the numerical one for $ N=10 $.

\section{Autocorrelator at $ T=\infty $}%
	\begin{figure}
		\centering
		\includegraphics[width=1.0\linewidth]{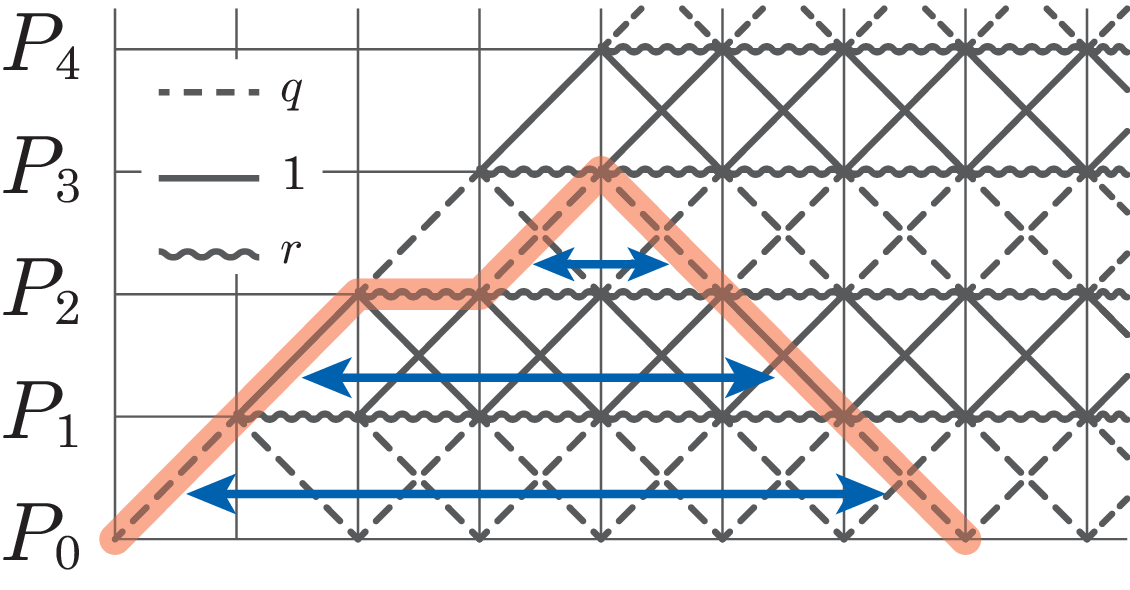}
		\caption{An example of the weighted Riordan paths. Each path is from $ (0,0) $ to $ (n,0) $ with up ($ \nearrow $), down ($ \searrow $), and horizontal ($ \rightarrow $) steps, never going below the bottom line nor containing horizontal steps on it. 
	Dashed (wavy horizontal) steps are endowed with a weight $q$ ($r$), which yields the weight of each Riordan path.
For instance, the red path from $ (0,0) $ to $ (7,0) $ has a weight $ q^4 r $. The sign of a weight is plus (respectively, minus) if 
%(\# of pairs of matching up steps and down steps 
(number of matched up-down pairs depicted by blue double-headed arrows) 
$+$ 
(number of horizontal steps) is even (respectively, odd).}
		\label{fig:autocorrelator_with_dissipation}
	\end{figure}In this section, we study the `infinite temperature' autocorrelator of the edge spin
	\begin{align}
		C_\infty(t)\coloneqq\langle\sigma_1^z(t)\sigma_1^z(0)\rangle_{T=\infty}=\dfrac{1}{2^N}\tr(\mathrm{e}^{t\mathcal{L}^\ast}\qty[\sigma_1^z]\sigma_1^z),\label{eq:autocorrelator}
	\end{align}
	where $ \mathcal{L}^\ast $ is the adjoint operator of the Liouvillian, which describes the time evolution of an operator $ X $ as follows:
	\begin{align}
	\dv{t}X(t)&=\mathcal{L}^\ast[X(t)]\notag\\
	&\coloneqq\mathrm{i}[H,X(t)]+\sum_{i}\qty(L_i^\dagger X(t)L_i-\dfrac{1}{2}\qty{L_i^\dagger L_i,X(t)}).
	\end{align}
	In other words, it corresponds to the Heisenberg picture of the open quantum system. A fundamental motivation in quantum engineering is for localized degrees of freedom to maintain coherence over long times. In Ref.~\cite{Kemp2017}, this quantity for the transverse-field Ising 
%chain 
and XYZ chains without dissipation (i.e., closed system) has been studied as the witness of long coherence times for edge spins. The autocorrelator for the dissipative transverse-field Ising model %has also been 
was also studied in Ref.~\cite{Vasiloiu2018}. However, very few exact results are available for the time evolution of physical quantities~\cite{Brandt1976,Foss-Feig2017}. Here, we study $ C_\infty(t) $ for our model with $ N=\infty $. Let us consider how the adjoint Liouvillian acts on $ \sigma_1^z $. For notational simplicity, we set $ J_x=q/2,\, J_y=1/2 $ and $ r\coloneqq 4\gamma $. In this case, one finds
\begin{align}
   \mathcal{L}^\ast[P_0]&=q P_1,
\end{align}
and for $ n\ge 1 $, 
\begin{align}
\mathcal{L}^\ast[P_n]&=
			\begin{dcases}
				-qP_{n-1}-rP_n-P_{n+1}&(n\text{: odd})\\
				P_{n-1}-rP_n+qP_{n+1}&(n\text{: even})
			\end{dcases},
		\label{eq:ad_Liouvillian_step_with_dissipation}
\end{align}
where 
\begin{align}
	P_0&=\sigma_1^z,\notag\\
		P_n&=
		\begin{dcases}
			-\sigma_1^y\qty(\prod_{i=2}^{n}\sigma_i^z)\sigma_{n+1}^x& (n\text{: odd})\\
			-\sigma_1^y\qty(\prod_{i=2}^{n}\sigma_i^z)\sigma_{n+1}^y& (n\ge 2 ~\text{and even})
		\end{dcases}.
\end{align}
%	\begin{align}
%		&\hspace{-2.5em}\begin{aligned}
%			\mathcal{L}^\ast[P_0]&=q P_1,\\
%			\text{and for }n\ge 1,&\\
%			\mathcal{L}^\ast[P_n]&=
%			\begin{dcases}
%				-qP_{n-1}-rP_n-P_{n+1}&(n\text{: odd})\\
%				P_{n-1}-rP_n+qP_{n+1}&(n\text{: even})
%			\end{dcases}
%		\end{aligned},\label{eq:ad_Liouvillian_step_with_dissipation}\\
%		\text{where}&\notag\\
%		P_0&=\sigma_1^z,\notag\\
%		P_n&=
%		\begin{dcases}
%			-\sigma_1^y\qty(\prod_{i=2}^{n}\sigma_i^z)\sigma_{n+1}^x& (n\text{: odd})\\
%			-\sigma_1^y\qty(\prod_{i=2}^{n}\sigma_i^z)\sigma_{n+1}^y& (n\text{: even and }n\ge 2).
%		\end{dcases}
%	\end{align}
	It is important to note that $ P_n $'s are Hermitian and form an orthonormal set, i.e.,
	\begin{align*}
		\langle\!\langle P_i,\, P_j\rangle\!\rangle\coloneqq \dfrac{1}{2^N}\tr(P_i^\dagger P_j)=\delta_{ij}.
	\end{align*}
	The inner product $ \langle\!\langle\cdot ,\cdot \rangle\!\rangle $ for matrices is called the Hilbert-Schmidt inner product, %Using it, 
with which Eq. (\ref{eq:autocorrelator}) takes the form
	\begin{align}
		C_\infty(t)=\sum_{n=0}^\infty \dfrac{t^n}{n!}\left\langle\langle {\mathcal{L}^\ast}^n\qty[P_0],\, P_0\right\rangle\rangle.\label{eq:autocorrelator_recast}
	\end{align}
	%Then, 

\begin{figure*}
	\centering
	\includegraphics[width=0.9\linewidth]{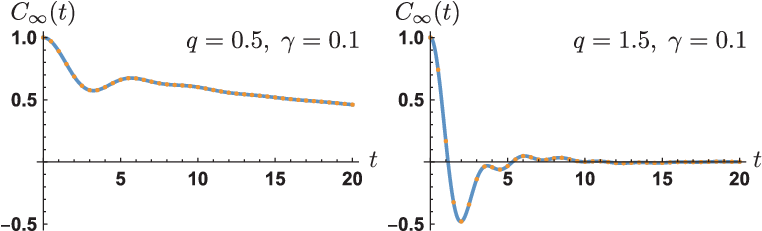}
	\caption{(Color online) Autocorrelator $ C_\infty(t) $ with $ \gamma=0.1 $ in (left) the topological regime $ q=0.5 $ and (right) the trivial regime $ q=1.5 $. Blue lines are obtained from the analytical result Eq. (\ref{eq:C_infty_results}) in Appendix \ref{app:exact_formula_of_the_autocorrelator_with_dissipation}, while orange dots are numerical results with $ N=100 $ by the Runge-Kutta method.}
	\label{fig:autocorrelator_plot}
\end{figure*}

Now we compute $ C_\infty(t) $ by considering the so called ``Riordan paths'' \cite{Chen2008,Cohen2016} (Motzkin paths \cite{Stanley1999} with no horizontal steps at the bottom line) weighted through $ q $ and $ r $. (see Fig. \ref{fig:autocorrelator_with_dissipation}).
To this end, it is useful to consider 
%Next, we obtain 
the generating function $ F(z;q,r) $ of the weighted Riordan paths, which can be obtained by the so called ``Kernel method'' \cite{Prodinger2004}. 
The autocorrelator in terms of $ F(z;q,r) $ reads
	\begin{align}
	C_\infty(t)&=\sum_{n=0}^{\infty}\dfrac{t^n}{n!}\qty[z^n]F(z;q,r),
	\end{align}
	where %$ \qty[z^n]F(z;q,r) $ 
$ \qty[z^n] f(z) $ denotes the coefficient of $ z^n $ in $f(z)$. 
%It can be recast by the contour integration as
This can be rewritten as a contour integral 
	\begin{align}
	C_\infty(t)&=\sum_{n=0}^{\infty}\dfrac{t^n}{n!}\oint \dfrac{\dd z}{2\pi\mathrm{i}}\dfrac{F(z; q,r)}{z^{n+1}}\\
	&=\dfrac{1}{2\pi\mathrm{i}}\oint \dfrac{F(1/w;q,r)}{w}\mathrm{e}^{tw}\, \dd w,
	\end{align}
	where $ w=1/z $. Here we have chosen the contour in the $ z $-plane so that it surrounds the origin and is sufficiently small. As a result, the contour in the $ w $-plane is sufficiently large.  
%we have chosen the contour of $ z $ sufficiently small around the origin, hence, the contour of $ w=1/z $ sufficiently large. 
The final explicit expression for $C_\infty(t)$ is cumbersome and is shown in Appendix \ref{app:exact_formula_of_the_autocorrelator_with_dissipation}. The analytic and finite-size numerical results of $ C_\infty(t) $ with $ \gamma=0.1 $ in the topological/trivial regime are shown in Fig. \ref{fig:autocorrelator_plot}.

From the exact formula, we can derive the inverse of the decay time $ \tau $ of $ C_\infty(t) $ ($ C_\infty(t)\stackrel{t\to\infty}{\sim}\mathrm{e}^{-t/\tau} $) with nonzero $ r $: for $ 0 < q < 1 $, one has
	\begin{align}
		\tau^{-1}&=-\eta_+(q,r),
	\end{align}
and for $ q > 1$,
	\begin{align}
		\tau^{-1}&=
		\begin{dcases}
			r&(0<r\le 1)\\
			\dfrac{1+r^2}{2r}&\qty(1\le r\le q+\sqrt{q^2-1})\\
			-\eta_+(q,r)&\qty(r\ge q+\sqrt{q^2-1})
		\end{dcases},
	\end{align}
%	\begin{align}
%		\text{for $ 0<q<1 $,}\quad \tau^{-1}&=-\eta_+(q,r)\\
%		\text{for $ q>1 $,}\quad \tau^{-1}&=
%		\begin{dcases}
%			r&(0<r<1)\\
%			\dfrac{1+r^2}{2r}&\qty(1<r<q+\sqrt{q^2-1})\\
%			-\eta_+(q,r)&\qty(r>q+\sqrt{q^2-1})
%		\end{dcases},
%	\end{align}
where $ \eta_+(q,r)\coloneqq (-1-r^2+\sqrt{(1+r^2)^2-4q^2r^2})/(2r) $. In particular, we find that the decay is suppressed in the topological regime ($ 0<q<1 $), although $ C_\infty(t) $ goes to $ 0 $ as $ t\to \infty $ in both regimes. One might note that $ C_\infty (t) $ behaves differently from the Liouvillian gap $ g $; $ g $ has a cusp for every, i.e., topological, critical, and trivial regime, while $ C_\infty (t) $ has non-analytical points only for critical and trivial regime corresponding to $ q \ge 1 $. There is, however, no contradiction between them. This is because $ g $ is determined by the NESSs and the first decay mode, while $ C_\infty (t) $ is obtained only from the completely mixed state, i.e., the mixture of the two NESSs.

\section{Conclusions}%
	We have studied the one-dimensional quantum compass model with dephasing noise and obtained the exact steady states. 
We showed that the model can be mapped to the non-Hermitian Kitaev model on a ladder, which is solvable by representing the spins in terms of Majorana fermions. 
%model and with his technique we have studied the Liouvillian gap.
This technique allows us to study the Liouvillian gap exactly. In particular, in the critical case where $J_x=J_y$, the gap in the thermodynamic limit is obtained analytically.    
We have also studied the autocorrelator of the edge spin and obtained 
%the exact formula of the autocorrelator 
its exact formula %with the aid of 
using the technique of combinatorics. 

\begin{acknowledgments}
	We would like to thank T. Prosen and M. \v{Z}nidari\v{c} for valuable comments. H.K. was supported in part by JSPS KAKENHI Grant No. JP18H04478 and No. JP18K03445. N.S. acknowledges support of the Materials Education program for the future leaders in Research, Industry, and Technology (MERIT). 
\end{acknowledgments}

\appendix

\section{\label{app:Lindbladian_non-Hermitian}MAPPING OF THE LIOUVILLIAN TO THE NON-HERMITIAN HAMILTONIAN}
	In this Appendix, we describe the identification of the Liouvillian with a non-Hermitian Hamiltonian in detail. Let $ A $ be a linear operator on Hilbert space $ \mathscr{H} $ (i.e., $ A\in \mathrm{End}_\mathbb{C}(\mathscr{H}) $). Assuming $ \dim(\mathscr{H})=\mathcal{N} $, there is a complete orthonormal basis $ \qty{\ket{\phi_i}}_{i=1}^\mathcal{N} $ of $ \mathscr{H} $, and $ A $ can be regarded as an $ \mathcal{N}\times \mathcal{N} $ matrix
	\begin{equation}
		A=\sum_{i,j=1}^{\mathcal{N}} \mathsf{A}_{ij}\ket{\phi_i}\bra{\phi_j},
	\end{equation}
	where a sans-serif style $ \mathsf{A}\in M_{\mathcal{N}\times \mathcal{N}}(\mathbb{C}) $ and its element $ \mathsf{A}_{ij} $ is just a c-number. Now, we 
%can define 
introduce a new space $ \mathrm{Ket}\otimes\mathrm{Bra} $ of dimension $ \mathcal{N}^2 $ by the following linear map $ F $:
	\begin{equation}
		\vcenter{
			\xymatrix@C=12pt@R=0.2pt{
				F:\hspace{-1em}&\mathrm{End}_\mathbb{C}(\mathscr{H})\ar[r]&\mathrm{Ket}\otimes\mathrm{Bra}\\
				&\rotatebox{90}{$ \in $}&\rotatebox{90}{$ \in $}\\
				&\ket{\phi_i}\bra{\phi_j}\ar@{|->}[r]&|\phi_i,\phi_j \rangle\rangle
			}
		}
	\end{equation}
	Note that $ F $ depends on the choice of the basis $ \qty{\ket{\phi_i}}_{i=1}^\mathcal{N} $, but after fixing the basis, $ F $ is an isomorphism, i.e., $ \mathrm{End}_\mathbb{C}(\mathscr{H}) $ and $ \mathrm{Ket}\otimes\mathrm{Bra} $ are in one-to-one correspondence. 

	Let us consider how superoperators ($ \in\mathrm{End}_\mathbb{C}(\mathrm{End}_\mathbb{C}(\mathscr{H}))  $) look like in the $ \mathrm{Ket}\otimes\mathrm{Bra} $ space. All superoperators in the Liouvillian $ \mathcal{L}[\rho] $ have the form
	\begin{equation}
		\rho\mapsto A\rho B.
	\end{equation}
	This map is rewritten as
	\begin{equation}
		\sum_{i,j}\mathsf{R}_{ij}\ket{\phi_i}\bra{\phi_j}\mapsto \sum_{i,j}\qty(\mathsf{A}\mathsf{R}\mathsf{B})_{ij}\ket{\phi_i}\bra{\phi_j}
	\end{equation}
	for $ A=\sum_{i,j} \mathsf{A}_{ij}\ket{\phi_i}\bra{\phi_j} $, $ B=\sum_{i,j} \mathsf{B}_{ij}\ket{\phi_i}\bra{\phi_j} $ and $ \rho=\sum_{i,j}\mathsf{R}_{ij}\ket{\phi_i}\bra{\phi_j} $. In the $ \mathrm{Ket}\otimes\mathrm{Bra} $ space, the superoperator can be seen as the following map:
	\begin{widetext}
		\begin{align}
			\sum_{i,j}\mathsf{R}_{ij}|\phi_i,\phi_j\rangle\rangle&\mapsto \sum_{i,j}\qty(\mathsf{A}\mathsf{R}\mathsf{B})_{ij}|\phi_i,\phi_j\rangle\rangle\nonumber\\
			&=\qty[\qty(\sum_{i,k}\mathsf{A}_{ik}|\phi_i\rangle\rangle\langle\langle\phi_k|)_\mathrm{Ket}\otimes\qty(\sum_{j,l}\mathsf{B}_{jl}^\mathrm{T}|\phi_j\rangle\rangle\langle\langle\phi_l|)_\mathrm{Bra}]\sum_{m,n}\mathsf{R}_{mn}|\phi_m,\phi_n\rangle\rangle.
		\end{align}
	\end{widetext}
	Therefore, this superoperator $ \rho\mapsto A\rho B $ can be thought of as the tensor product of two matrix $ \mathsf{A}\otimes\mathsf{B}^\mathrm{T} $. Note that this matrix is basis-dependent because it is not $ \mathsf{A}\otimes \mathsf{B}^\dagger $. Then, one can identify the Liouvillian $ \mathcal{L} $ in Eq. (\ref{eq:Lindblad}) as
	\begin{align}
		\mathcal{L}&\cong-\mathrm{i}H\otimes\mathds{1}+\mathrm{i}\mathds{1}\otimes H^\mathrm{T}\nonumber\\
		&\quad+\sum_{i}\qty(L_i\otimes L_i^\ast-\dfrac{1}{2}L_i^\dagger L_i\otimes\mathds{1}-\dfrac{1}{2}\mathds{1}\otimes L_i^\mathrm{T}L_i^\ast).\label{eq;Lindbladian_maps_to_non-Hermitian}
	\end{align}
	Here, we do not distinguish operators in italic with matrices in sans-serif. For Eq. (\ref{eq:XY_ladder_Hamiltonian}), transpose or conjugate in Eq. (\ref{eq;Lindbladian_maps_to_non-Hermitian}) does not matter if we choose a basis which diagonalizes $ \sigma_i^z $'s.

\section{\label{app:non-Hermitian_quadratic_form_of_fermions}NON-HERMITIAN QUADRATIC FORM OF FERMIONS}
	In a usual closed free-fermion system, we can generally write the Hamiltonian as follows:
	\begin{equation}
		H=\vb*{f}^\dagger\mathsf{A} \vb*{f},\label{eq:non-Hermitian_quadratic_form_of_fermions_1}
	\end{equation}
	where $ \vb*{f}=(f_1,\dots, f_N)^\mathrm{T},\, \vb*{f}^\dagger=(f_1^\dagger,\dots, f_N^\dagger) $, and $ \mathsf{A} $ is an $ N\times N $ Hermitian matrix. In general, all eigenvalues of a Hermitian matrix are real, and eigenvectors form an orthonormal basis (after normalization and orthogonalization in degenerate spaces). In other words, $ \mathsf{A} $ can be diagonalized by a unitary matrix $ \mathsf{U} $, and the Hamiltonian is rewritten as
	\begin{equation}
		H=\vb*{f}^\dagger \mathsf{U}\, \mathrm{diag}(a_1,\dots,a_N)\, \mathsf{U}^\dagger\vb*{f},
	\end{equation}
	where $ a_i $ $ (i=1,\dots, N) $ are the eigenvalues of $ \mathsf{A} $. Then, we can define new operators $ \vb*{f}'\coloneqq \mathsf{U}^\dagger\vb*{f} $, and it is easily verified that $ \vb*{f}' $ also satisfy anticommutation relations. Therefore, we obtain
	\begin{equation}
		H=\sum_{i=1}^N a_i {f'_i}^\dagger f'_i,\label{eq:non-Hermitian_quadratic_form_of_fermions_2_diagonalization}
	\end{equation}
	and all eigenvalues of $ H $ are obtained by $ E=\sum_i a_in_i $ with arbitrary choice of each $ n_i=0\text{ or } 1 $.
	
	However, some care must be taken when $ \mathsf{A} $ is non-Hermitian. First, eigenvalues of $ \mathsf{A} $ are not necessarily real. Second, eigenvectors with different eigenvalues are in general not orthogonal. Third, left and right eigenvectors with the same eigenvalue are in general not a Hermitian-conjugate to each other. We briefly explain the general prescription for treating non-Hermitian matrices according to Ref. \cite{Brody2014}. A similar discussion can be found in Ref. \cite{Sternheim1972}.

\medskip
	
	Let $ \mathsf{A} $ be a non-Hermitian and non-degenerate $ N\times N $ matrix. Remember that if a matrix is non-degenerate, then it is diagonalizable. \footnote{We can easily generalize the following discussion to the case where $ \mathsf{A} $ is degenerate but diagonalizable. However, some degenerate matrices cannot be diagonalized, and such situations are sometimes called ``coalescence'' instead of ``degeneracy''. We do not consider ``coalescence'' here. See, e.g., Ref. \cite{Heiss2012} for more details about coalescence.} Then, $ \mathsf{A} $ has left/right eigenvectors $ \{\vb*{l}_i^\dagger\}_{i=1}^N $ and $ \{\vb*{r}_j\}_{j=1}^N $ which satisfy
	\begin{align}
		\vb*{l}_i^\dagger\mathsf{A}&=\lambda_i\vb*{l}_i^\dagger,\\
		\mathsf{A}\vb*{r}_j&=\xi_j\vb*{r}_j.
	\end{align}
	($ \vb*{l}_i $ and $ \vb*{r}_j $ are column vectors.) From these, we obtain
	\begin{gather}
		\lambda_i\langle \vb*{l}_i,\, \vb*{r}_j\rangle=\vb*{l}_i^\dagger \mathsf{A} \vb*{r}_j=\xi_j \langle \vb*{l}_i,\, \vb*{r}_j\rangle\nonumber\\
		\therefore\quad (\lambda_i-\xi_j)\langle \vb*{l}_i,\, \vb*{r}_j\rangle=0,
	\end{gather}
	where $ \langle \vb*{l}_i,\, \vb*{r}_j\rangle\coloneqq \vb*{l}_i^\dagger \vb*{r}_j $ is the standard inner product of $ \mathbb{C}^N $. Because $ \mathsf{A} $ is diagonalizable, each of $ \qty{\vb*{l}_i}_{i=1}^N $ and $ \qty{\vb*{r}_j}_{j=1}^N $ is a basis of $ \mathbb{C}^N $. Therefore, for each $ i $, at least one $ j $ satisfies $ \langle \vb*{l}_i,\, \vb*{r}_j\rangle\ne 0 $. Then, we can assume $ \lambda_i=\xi_i $ after relabeling $ j $'s. It follows that $ \langle \vb*{l}_i,\, \vb*{r}_j\rangle=0 $ if $ i\ne j $ and $ \langle \vb*{l}_i,\, \vb*{r}_i\rangle\ne 0 $. Therefore, after ``normalization'' of $ \qty{\vb*{l}_i}_{i=1}^N $ and/or $ \qty{\vb*{r}_j}_{j=1}^N $ so as to satisfy $ \langle \vb*{l}_i,\, \vb*{r}_i\rangle=1 $, we obtain
	\begin{equation}
		\langle \vb*{l}_i,\, \vb*{r}_j\rangle=\delta_{ij}.
	\end{equation}
	Then, we define
	\begin{equation}
		\mathsf{V}\coloneqq \qty(\vb*{r}_1 \dots \vb*{r}_N) \quad \qty(\iff \mathsf{V}^{-1}=
		\begin{pmatrix}
			\vb*{l}_1^\dagger\\
			\vdots\\
			\vb*{l}_N^\dagger
		\end{pmatrix}
		),
	\end{equation}
	and $ \mathsf{A} $ is diagonalized as
	\begin{equation}
		\mathsf{V}^{-1}\mathsf{A}\mathsf{V}=\mathrm{diag}(\lambda_1,\dots, \lambda_N).
	\end{equation}
	
	Now, let us return to Eq. (\ref{eq:non-Hermitian_quadratic_form_of_fermions_1}) with non-Hermitian $ \mathsf{A} $. By diagonalizing $ \mathsf{A} $ as above, we obtain
	\begin{equation}
		H=\vb*{f}^\dagger\mathsf{V}\, \mathrm{diag}(\lambda_1,\dots, \lambda_N)\, \mathsf{V}^{-1}\vb*{f}
	\end{equation}
	and we define
	
	\begin{equation}
		a_i\coloneqq\vb*{l}_i^\dagger \vb*{f},\quad b_i^\dagger\coloneqq \vb*{f}^\dagger\vb*{r}_i,
	\end{equation}
	then, the Hamiltonian can be written in the form similar to Eq. (\ref{eq:non-Hermitian_quadratic_form_of_fermions_2_diagonalization}),
	\begin{equation}
		H=\sum_{i=1}^N \lambda_i b_i^\dagger a_i.
	\end{equation}
	One can easily verify the following anticommutation relations
	\begin{equation}
		\qty{a_i,b_j^\dagger}=\delta_{ij},\quad \qty{a_i,a_j}=\qty{b_i^\dagger,b_j^\dagger}=0.
	\end{equation}
	Therefore, $ a_i $ and $ b_i^\dagger $ can be seen as an annihilation and an creation operator of new fermions, respectively, although $ a_i^\dagger $ is not equal to $ b_i^\dagger $ unlike Hermitian cases \cite{Bilstein1998}. From these anticommutation relations, it follows that $ b_i^\dagger a_i $ has eigenvalues $ 0 $ and $ 1 $, although it is not Hermitian. Then we obtain all eigenvalues of $ H $ as $ E=\sum_i \lambda_in_i $ with arbitrary choice of each $ n_i=0\text{ or } 1 $.
	
\section{\label{app:from_Kitaev_to_SSH}FROM KITAEV LADDER TO SSH MODEL}
	By Kitaev's mapping, a non-Hermitian Hamiltonian $ \mathcal{H} $ can be seen as a model of free Majorana fermions in a static $ \mathbb{Z}_2 $ gauge field. Introducing Majorana fermion operators as $ \sigma_j^\alpha\to \mathrm{i} b_j^\alpha c_j $ and $ \tau_j^\alpha\to \mathrm{i}\tilde{b}_j^\alpha d_j $ ($ i=1,\dots, N,\, \alpha=x,y,z $), we obtain
	\begin{widetext}
		\begin{align}
		\mathcal{H}&=J_x\sum_{j=1}^{N/2}\qty[(\mathrm{i}b_{2j-1}^xb_{2j}^x)(\mathrm{i}c_{2j-1}c_{2j})-(\mathrm{i}\tilde{b}_{2j-1}^x\tilde{b}_{2j}^x)(\mathrm{i}d_{2j-1}d_{2j})]+J_y\sum_{j=1}^{N/2-1}\qty[(\mathrm{i}b_{2j}^yb_{2j+1}^y)(\mathrm{i}c_{2j}c_{2j+1})-(\mathrm{i}\tilde{b}_{2j}^y\tilde{b}_{2j+1}^y)(\mathrm{i}d_{2j}d_{2j+1})]\notag\\
		&\hspace{1em}-\mathrm{i}\gamma \sum_{i=1}^{N}(\mathrm{i}b_i^z\tilde{b}_i^z)(\mathrm{i}c_id_i),
		\end{align}
	\end{widetext}
%	The $ \mathbb{Z}_2 $ gauge degree of freedom is equal to the number of plaquettes in the system. 
where the operators, $ \mathrm{i}b_{2j-1}^xb_{2j}^x $, $ \mathrm{i}b_{2j}^yb_{2j+1}^y $, $ \mathrm{i}\tilde{b}_{2j-1}^x\tilde{b}_{2j}^x $, $ \mathrm{i}\tilde{b}_{2j}^y\tilde{b}_{2j+1}^y $, and $ \mathrm{i}b_i^z\tilde{b}_i^z $ commute with the Hamiltonian and their eigenvalues are  $\pm 1 $. Therefore the Hilbert space splits into sectors labeled by the eigenvalues of these operators. One can define the flux  through each plaquette by the eigenvalue of the product of $ b $ and $ {\tilde b} $ operators around it. The spectrum of the Hamiltonian $ {\cal H} $ depends only on the set of fluxes. Thus we can fix all but $ N-1 $ signs of the links, as we have $ N-1 $ plaquettes in our model. We fix them as
	\begin{align}
		\mathrm{i}b_{2j-1}^xb_{2j}^x=\mathrm{i}b_{2j}^yb_{2j+1}^y=-1,\quad \mathrm{i}\tilde{b}_{2j-1}^x\tilde{b}_{2j}^x=\mathrm{i}\tilde{b}_{2j}^y\tilde{b}_{2j+1}^y=+1
	\end{align}
	and define $ \mu_i=-\mathrm{i}b_i^z\tilde{b}_i^z $ to recast $ \mathcal{H} $ as
	\begin{align}
		\mathcal{H}&=-J_x\sum_{j=1}^{N/2}\qty[(\mathrm{i}c_{2j-1}c_{2j})+(\mathrm{i}d_{2j-1}d_{2j})]\notag\\
		&\hspace{1em}-J_y\sum_{j=1}^{N/2-1}\qty[(\mathrm{i}c_{2j}c_{2j+1})+(\mathrm{i}d_{2j}d_{2j+1})]\notag\\
		&\hspace{1em}+\mathrm{i}\gamma \sum_{i=1}^{N}\mu_i(\mathrm{i}c_id_i).
	\end{align}
	Next, we define complex fermions $ f_i $ and $ f_i^\dagger $, 
%which is made 
consisting of two Majorana fermions $ c_i $ and $ d_i $:
	\begin{align}
		f_i\coloneqq \dfrac{c_i+\mathrm{i}d_i}{2},\quad f_i^\dagger\coloneqq \dfrac{c_i-\mathrm{i}d_i}{2}.
	\end{align}
	It is easy to verify that they satisfy the anticommutation relations
	\begin{align}
		\qty{f_i,f_j^\dagger}=\delta_{ij}, \quad
     \qty{f_i, f_j } = \qty{f^\dagger_i, f^\dagger_j} = 0.
	\end{align}
	We then have
	\begin{align}
		\mathcal{H}&=2\mathrm{i}J_x\sum_{i=1}^{N/2}\qty(f_{2j}^\dagger f_{2j-1}-f_{2j-1}^\dagger f_{2j})\notag\\
		&\hspace{1em}+2\mathrm{i}J_y\sum_{i=1}^{N/2-1}\qty(f_{2j+1}^\dagger f_{2j}-f_{2j}^\dagger f_{2j+1})\notag\\
		&\hspace{1em}+\mathrm{i}\gamma\sum_{i=1}^N \mu_i\qty(2f_i^\dagger f_i-1).
	\end{align}
	After the unitary transformation $ f_j\to \mathrm{e}^{\mathrm{i}(\pi/2)j}f_j $, we obtain Eq. (\ref{eq:non-Hermitian_SSH_1}).
	
\section{\label{app:the_first_decay_modes_configurations}THE FIRST DECAY MODES' CONFIGURATIONS}
	In Fig. \ref{fig:lindblad-kitaevgvsgap}, we show for each phase only one example of the configurations where the first decay mode lives, but numerical calculation reveals that there are other such configurations. Tab. \ref{tab:first_decay_mode} shows the numerical results of 
%every such a configuration 
all such configurations up to symmetries mentioned in the main text.
	\begin{table*}
		\caption{The chemical-potential configurations where the first decay modes live.}
		\label{tab:first_decay_mode}
		\begin{ruledtabular}
			\begin{tabular}{c|c|c|c}
				&very small-$ \gamma $ phase&phase I&phase I\hspace{-0.1em}I\\
				\hline
				topological&
				$ \begin{cases}
				\mu_i=-1&i=1,N\\
				\mu_i=+1&\text{otherwise}
				\end{cases} $
				&
				$ \begin{cases}
					\mu_i=-1&i=1\\
					\mu_i=+1&\text{otherwise}
				\end{cases} $
				&
				$ \begin{cases}
					\mu_i=-1&i=1,2,3\\
					\mu_i=+1&\text{otherwise}
				\end{cases} $\\
				\hline
				critical&\multicolumn{2}{c|}{$ \exists ! i\text{ s.t. }\mu_i=-1,\text{ and }\mu_j=+1\text{ for }j\ne i $}&
				$ \begin{cases}
					\mu_i=-1&i=1,2\\
					\mu_i=+1&\text{otherwise}
				\end{cases} $\\
				\hline
				trivial&\multicolumn{2}{c|}{$ \exists ! i\text{ s.t. }\mu_i=-1,\text{ and }\mu_j=+1\text{ for }j\ne i $}&
				$ \begin{cases}
					\mu_i=-1&i=1,2,3,4\\
					\mu_i=+1&\text{otherwise}
				\end{cases} $
			\end{tabular}
		\end{ruledtabular}
	\end{table*}
\section{\label{app:derivation_of_Liouvillian_gap}DERIVATION OF EQ. (\ref{eq:exact_Liouvillian_gap})}
	We call the pattern of $ \vb*{\mu} $ which satisfies ``$ \mu_1=-1,\text{ otherwise }\mu_i=+1 $'' ``pattern 1,'' and that which satisfies ``$ \mu_1=\mu_2=-1,\text{ otherwise }\mu_i=+1 $'' ``pattern 2.'' $ \mathsf{A}_{1} $ (respectively, $ \mathsf{A}_{2} $) denotes the matrix $ \mathsf{A} $ with the pattern 1 (respectively, pattern 2). The (unnormalized) eigenstates $ \vb*{v}=(v_1,v_2,\dots, v_N) $ of $ \mathsf{A}_{1} $ whose eigenvalues have negative imaginary parts are obtained by the following ansatz
	\begin{equation}
		v_{2n-1}=\alpha^{n-1}, \quad v_{2n}=\mathrm{i}\beta\alpha^{n-1}\quad (n=1,\dots,N/2),
	\end{equation}
	where $ \alpha,\beta\in \mathbb{C} $ and $ \abs{\alpha}<1 $. Letting $ \lambda $ be the eigenvalue of $ \mathsf{A}_1 $ corresponding to $ \vb*{v} $, we obtain the following conditions for this ansatz
	\begin{align}
		\text{left boundary:}&& &-2\mathrm{i}\gamma + 2\mathrm{i}J_x\beta=\lambda_1,\\
		\text{bulk:}&& &
		\begin{cases}
			2J_x-2\gamma\beta+2J_y\alpha=\mathrm{i}\beta\lambda_1\\
			2\mathrm{i}J_y\beta+2\mathrm{i}\gamma\alpha+2\mathrm{i}J_x\alpha\beta=\alpha\lambda_1
		\end{cases}.
	\end{align}
	Here, we neglect the right boundary condition that is justified in the thermodynamic limit. The solution of $ \lambda $ is
	\begin{equation}
		\lambda_1=-\dfrac{\mathrm{i}}{2\gamma}\qty[-J_y^2+\sqrt{8\gamma^2(2\gamma^2+J_y^2-2J_x^2)+J_y^4}].
	\end{equation}
	In the critical case of $ J_x=J_y=1 $, $ \lambda $ has negative imaginary part when $ \gamma>1/\sqrt{2} $, and the gap for pattern 1 [i.e., the argument in parentheses in Eq. (\ref{eq:Liouvillian_gap_simplified}) for pattern 1] reads
	\begin{equation}
		g_1=
		\begin{dcases}
			2\gamma&\qty(0\le \gamma\le 1/\sqrt{2})\\
			\dfrac{1}{\gamma}&(1/\sqrt{2}\le\gamma)
		\end{dcases}.
	\end{equation}
	In a similar way, we obtain the localized solution for pattern 2 by the ansatz
	\begin{equation}
		\begin{cases}
			v_1=1, v_2=\mathrm{i}\delta\beta, \\
			v_{2n-1}=\delta\alpha^{n-1}, v_{2n}=\mathrm{i}\delta\beta\alpha^{n-1}\; (n=2,\dots, N/2)
		\end{cases},
	\end{equation}
	and conditions
	\begin{align}
		\text{left boundary: }&& &
		\begin{cases}
			-2\mathrm{i}\gamma+2\mathrm{i}J_x\delta\beta=\lambda_2\\
		2J_x+2\gamma\delta\beta+2J_y\delta\alpha=\mathrm{i}\delta\beta\lambda_2
		\end{cases},\\
		\text{bulk: }&& &
		\begin{cases}
			2J_x-2\gamma\beta+2J_y\alpha=\mathrm{i}\beta\lambda_2\\
			2\mathrm{i}J_y\beta+2\mathrm{i}\gamma\alpha+2\mathrm{i}J_x\alpha\beta=\alpha\lambda_2
		\end{cases}.
	\end{align}
	From these conditions, we obtain in the critical case the gap for pattern 2 as
	\begin{widetext}
		\begin{equation}
			g_2=
			\begin{dcases}
				4\gamma&\qty(0\le\gamma\le\sqrt{(\sqrt{5}-1)/8})\\
				\frac{6^{1/3}\qty(9 \gamma^2+\sqrt{48 \gamma^6+81\gamma^4})^{2/3}-2\cdot 6^{2/3} \gamma^2}{3 \gamma(9 \gamma^2+\sqrt{48 \gamma^6+81 \gamma^4})^{1/3}}&\qty(\sqrt{(\sqrt{5}-1)/8}\le\gamma)
			\end{dcases}.
		\end{equation}
	\end{widetext}
	Finally, we obtain the global gap as
	\begin{widetext}
		\begin{align*}
			g&=\min(g_1,g_2)\\
			&=
			\begin{dcases}
				2\gamma&\qty(0\le \gamma\le\sqrt{(\sqrt{3}-1)/2})\\
				\frac{6^{1/3}\qty(9 \gamma^2+\sqrt{48 \gamma^6+81\gamma^4})^{2/3}-2\cdot 6^{2/3} \gamma^2}{3 \gamma(9 \gamma^2+\sqrt{48 \gamma^6+81 \gamma^4})^{1/3}}&\qty(\sqrt{(\sqrt{3}-1)/2}\le\gamma)
			\end{dcases}.
		\end{align*}
	\end{widetext}
	Therefore, the transition point $ \gamma_\mathrm{c} $ for the critical case is
	\begin{equation}
		\gamma_\mathrm{c}=\sqrt{\dfrac{\sqrt{3}-1}{2}}\simeq 0.605\dots
	\end{equation}
	One may guess that even for topological or trivial case, we can obtain the exact results for the gap in a similar way. However, it would be impossible to obtain the algebraic solutions because in these cases, we need to deal with equations of degree greater than four.
	
\section{\label{app:exact_formula_of_the_autocorrelator_with_dissipation}EXACT FORMULA FOR THE AUTOCORRELATOR WITH DISSIPATION}
	The generating function for the weighted Riordan path is obtained by the Kernel method~\cite{Prodinger2004} as
	\begin{widetext}
		\begin{align}
			F(z;q,r)\coloneqq \dfrac{-1+(1-q^2+r^2)z^2+\sqrt{\qty[\qty(1+q^2)z^2+(1+r z)^2]^2-4q^2 z^4}}{2z\qty[q^2r z^2+(1+r^2)z+r]}.
		\end{align}
	\end{widetext}
	Then, the autocorrelator is recast as
	\begin{widetext}
		\begin{align}
			C_\infty(t)&=\dfrac{1}{2\pi\mathrm{i}}\oint \dfrac{F(1/w; q,r)}{w}\mathrm{e}^{tw}\,\dd w\\
			&=\dfrac{1}{2\pi\mathrm{i}}\oint\dfrac{-w^2+1-q^2+r^2+\sqrt{\qty[(w+r)^2+(1+q)^2]\qty[(w+r)^2+(1-q)^2]}}{2r(w-\eta_{+}(q,r))(w-\eta_{-}(q,r))}\mathrm{e}^{tw}\,\dd w,\label{eq:autocorrelator_with_dissipation_contour}
		\end{align}
	\end{widetext}
	where
	\begin{align}
		\eta_\pm(q,r) &\coloneqq \dfrac{-1-r^2\pm\sqrt{(1+r^2)^2-4q^2r^2}}{2r}.
	\end{align}
	\begin{figure*}
		\centering
		\includegraphics[width=0.9\linewidth]{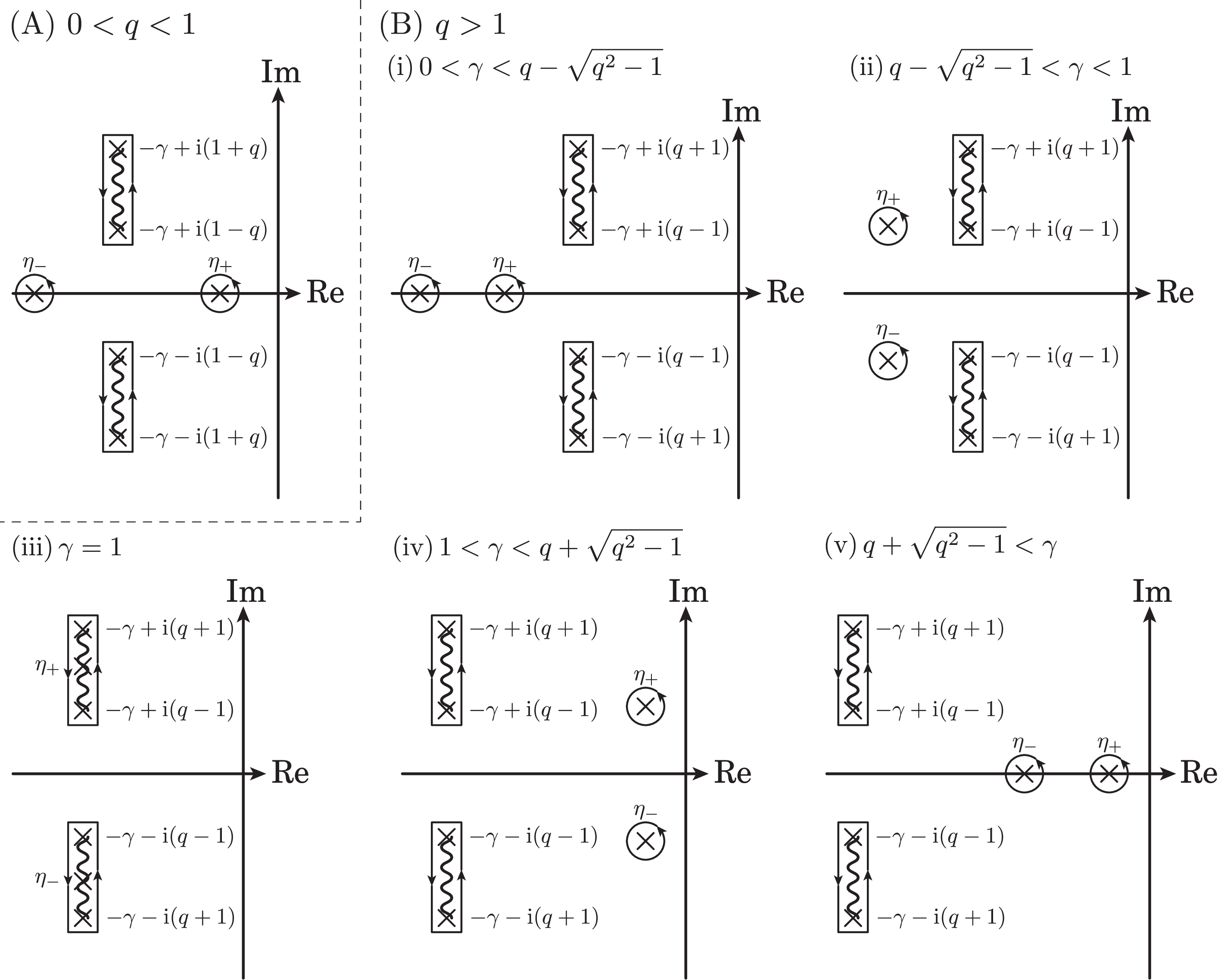}
		\caption{Integration contours of Eq. (\ref{eq:autocorrelator_with_dissipation_contour}). The crosses ($\times$) represent poles or branch points, while the wavy lines are branch cuts.}
		\label{fig:autocorrelator_with_dissipation_contour}
	\end{figure*}
	The contour of $ w $ is chosen as shown in Fig. \ref{fig:autocorrelator_with_dissipation_contour}, and the final results are
	\begin{widetext}
		\begin{align}
			C_\infty(t)&=
			\begin{dcases}
				\dfrac{-\eta_{+}^2+1-q^2+r^2}{r(\eta_{+}-\eta_{-})}\mathrm{e}^{\eta_{+} t}\\
				\quad+\dfrac{\mathrm{e}^{-r t}}{\pi r}\int_{1-q}^{1+q}f(y,q)\dfrac{(r^{-1}-r)y\cos(yt) +\qty[y^2-(r+\eta_{+})(r+\eta_{-})]\sin(yt)}{\qty[y^2+(r+\eta_{+})^2]\qty[y^2+(r+\eta_{-})^2]}\,\dd y&(0<q\le 1)\\
				\dfrac{\mathrm{e}^{-r t}}{\pi r}\int_{q-1}^{q+1}f(y,q)\dfrac{(r^{-1}-r)y\cos(yt) +\qty[y^2-(r+\eta_{+})(r+\eta_{-})]\sin(yt)}{\qty[y^2+(r+\eta_{+})^2]\qty[y^2+(r+\eta_{-})^2]}\,\dd y&(q\ge 1,\, 0\le r<1)\\
				\mathrm{e}^{-t}\cos(\sqrt{q^2-1}t)+\dfrac{\mathrm{e}^{-t}}{\pi}\int_{q-1}^{q+1}\dfrac{f(y,q)}{y^2-q^2+1}\sin(yt)\,\dd y&(q\ge 1,\, r=1)\\
				\dfrac{-\eta_{+}^2+1-q^2+r^2}{r(\eta_{+}-\eta_{-})}\mathrm{e}^{\eta_{+} t}-\dfrac{-\eta_{-}^2+1-q^2+r^2}{r(\eta_{+}-\eta_{-})}\mathrm{e}^{\eta_{-} t}\\
				\quad+\dfrac{\mathrm{e}^{-r t}}{\pi r}\int_{1-q}^{1+q}f(y,q) \dfrac{(r^{-1}-r)y\cos(yt) +\qty[y^2-(r+\eta_{+})(r+\eta_{-})]\sin(yt)}{\qty[y^2+(r+\eta_{+})^2]\qty[y^2+(r+\eta_{-})^2]}\,\dd y&(q\ge 1,\, r>1)
			\end{dcases},\label{eq:C_infty_results}
		\end{align}
	\end{widetext}
	where
	\begin{align}
		f(y,q)=\sqrt{\qty[(q+1)^2-y^2]\qty[y^2-(q-1)^2]}.
	\end{align}
	\pagebreak
	In general, it does not have a simpler form. However, in the absence of dissipation, i.e., when $ r=0 $, we have
	\begin{widetext}
		\begin{align}
			C_\infty(t;r=0)=
			\begin{dcases}
				1-q^2+\dfrac{1}{\pi}\int_{1-q}^{1+q}\dfrac{\sqrt{\qty[x^2-(1-q)^2]\qty[(1+q)^2-x^2]}}{x}\cos (xt) \, \dd x& (0<q\le 1)\\
				\dfrac{1}{\pi}\int_{q-1}^{q+1}\dfrac{\sqrt{\qty[x^2-(q-1)^2]\qty[(q+1)^2-x^2]}}{x}\cos (xt) \, \dd x& (q\ge 1).
			\end{dcases}.
		\end{align}
	\end{widetext}
	We have confirmed numerically that the integral goes to zero as $ t\to \infty $ in either case. Therefore, the autocorrelator is non-vanishing as $ t\to \infty $ in the topological phase, whereas vanishing in the trivial phase. Moreover, when $ q=1 $, it takes a simpler form:
	\begin{align}
		C_\infty(t;q=1,r=0)=\dfrac{J_1(2t)}{t},
	\end{align}
	where $ J_1 $ is the Bessel function of the first kind. From the asymptotic behavior of the Bessel function, we obtain that $ C_\infty (t) $ decays as $ \sim t^{-3/2} $ for large $ t $.
	
\bibliography{letter}% Produces the bibliography via BibTeX.

\end{document}